\documentclass[aps,prb,twocolumn,natbib,showpacs,floatfix,superscriptaddress, longbibliography]{revtex4-2}

\usepackage{graphicx}
\usepackage{amsmath,amssymb,bbold,bm,color}
\usepackage{float}
\usepackage{epstopdf}
\usepackage{hyperref}
\usepackage{color}
\usepackage{enumerate}
\usepackage{wasysym}
\usepackage{url}
\usepackage{dsfont}
\usepackage{braket}
\usepackage{comment}
\usepackage{siunitx}
\graphicspath{{figures/}}

\newcommand{\br}{\bm{r}}
\newcommand{\bR}{\bm{R}}
\newcommand{\bk}{\bm{k}}

\newcommand{\cT}{{\cal T}}

\newcommand{\cH}{{\cal H}}

\newcommand{\cF}{{\cal F}}

\newcommand{\tLambda}{\tilde{\Lambda}}
\newcommand{\pf}{{45^{\circ}}}

\hypersetup{colorlinks=true,linkcolor=blue,citecolor=blue,urlcolor=blue}

\begin{document}

\title{Incoherent tunneling and topological superconductivity in twisted cuprate bilayers}

\author{Rafael Haenel}
\affiliation{Department of Physics and Astronomy \& Stewart Blusson Quantum Matter Institute, University of British Columbia, Vancouver, BC V6T 1Z4, Canada}
\affiliation{Max Planck Institute for Solid State Research, 70569 Stuttgart, Germany}

\author{Tarun Tummuru}
\affiliation{Department of Physics and Astronomy \& Stewart Blusson Quantum Matter Institute, University of British Columbia, Vancouver, BC V6T 1Z4, Canada}
\affiliation{Department of Physics, University of Zurich, Winterthurerstrasse 190, Zurich 8057, Switzerland}

\author{Marcel Franz}
\affiliation{Department of Physics and Astronomy \& Stewart Blusson Quantum Matter Institute, University of British Columbia, Vancouver, BC V6T 1Z4, Canada}

\date{\today}

\begin{abstract}
   Twisting two monolayers of a high-$T_c$ cuprate superconductor can engender a chiral topological state with spontaneously broken time reversal symmetry $\mathcal{T}$. A crucial ingredient required for the emergence of a gapped topological phase is electron tunneling between the CuO$_2$ planes, whose explicit form (in an ideal clean sample) is dictated by the symmetry of the atomic orbitals. However, a large body of work on the interlayer transport in cuprates indicates importance of disorder-mediated incoherent tunneling which evades the symmetry constraints present in an idealized crystal. The latter arises even in the cleanest single-crystal samples through oxygen vacancies in layers separating the CuO$_2$ planes, introduced to achieve the hole doping necessary for superconductivity. Here we assess the influence of incoherent tunneling on the phase diagram of a twisted bilayer. We show that the model continues to support a fully gapped topological phase with broken $\mathcal{T}$, even in the limit of disorder-mediated interlayer coupling. Compared to the model with a constant, momentum conserving interlayer coupling, the extent of the topological phase around the 45$^\circ$ twist decreases with increasing incoherence, but remains robustly present for parameters likely relevant to Bi$_2$Sr$_2$CaCu$_2$O$_{8+\delta}$. 
\end{abstract}

\maketitle
 

\section{Introduction}

Twisted 2D van der Waals materials have emerged as an elegant platform to engineer and study correlated quantum phases with unprecedented experimental control \cite{Andrei2020, Novoselov2016, Cao2018, Balents2020, Wu2018, Regan2020}. At certain \textit{magic} angles, the electronic structure is deformed into exceedingly narrow bands in a moir\'e Brillouin zone. The small bandwidth yields pronounced correlation effects and generally makes multiple neighboring quantum phases experimentally accessible in a single fabricated device through application of electrostatic gating \cite{Cao2018}. 

Recently, a related paradigm of twisted bilayer structures was introduced that does not rely on engineering of correlated flat bands, but can produce interesting new phases by combining known non-trivial properties of constituent monolayers \cite{Can2021,volkov2021}. It was shown that two cuprate monolayers, stacked at a twist angle $\theta$, can give rise to a spontaneously time-reversal symmetry broken state by virtue of simple electron tunneling between the layers. Most notably, in a finite range of angles around the critical twist $\theta_c=45^\circ$, the ground state of an otherwise nodal $d$-wave superconductor becomes fully gapped and acquires a finite Chern number. At exactly $45^\circ$ the gapped phase persists up to the native critical temperature of the cuprate monolayer, thus furnishing the first known example of a high-$T_c$ topological superconductor. 

Pioneering experimental work on very thin twisted Bi$_2$Sr$_2$CaCu$_2$O$_{8+\delta}$ (BSCCO) flakes succeeded in fabricating bilayers at various twist angles \cite{Zhao2021}. Measurements of the interlayer Josephson current, Fraunhofer interference patterns and half-integer Shapiro steps in samples close to the $45^\circ$ are suggestive of a $\mathcal{T}$-broken phase \cite{Tummuru2022}. Strong twist angle dependence of the critical current has been reported elsewhere \cite{Lee2021}. 

It was later noted by Song, Zhang and Vishwanath \cite{Song2021} that twist angle and momentum dependence of the interlayer tunneling matrix element $g_{\bk}$, arising from the symmetry of the copper active orbitals, can play an important role in the emergence of the $\mathcal{T}$-broken phase. As originally argued by Xiang and Wheatley \cite{Xiang_1996}, the matrix element has the form
\begin{equation}\label{gk}
    g_{\bk}=g_0 \cos{2\theta}+g_1\mu_{\bk}(\theta/2)\mu_{\bk}(-\theta/2),
\end{equation}
which we generalized here to a twisted bilayer geometry following \cite{Song2021}. The $g_0$ term represents the direct tunneling between copper $d_{x^2-y^2}$ orbitals while the $g_1$ term describes the `oxygen-assisted' tunneling process with form factor $\mu_{\bk}(\theta)=[\cos{(R_\theta k_x)}-\cos{(R_\theta k_y)}]/2$ and $R_\theta$ the rotation matrix. Crucially, the part of $g_{\bk}$ that remains non-vanishing at $\theta=45^\circ$ contains the form factor $\mu_{\bk}$ that suppresses tunneling at the nodes of the $d$-wave order parameter. While spontaneous $\mathcal{T}$-breaking is still found to occur in this case, the ground state remains gapless and hence does not support the gapped topological phase with non-zero Chern number predicted in Ref.\,\cite{Can2021}.

In the present work we consider the twisted bilayer problem within a family of incoherent tunneling models \cite{Graf1993,Radtke1995,Radtke1996,Radtke_1997,Turlakov2001} in which the transfer of electrons between two adjacent CuO$_2$ layers is mediated by impurities that are inherently present in the otherwise inert `spacer' layers. Such incoherent tunneling models have been shown to yield better agreement with experimentally measured $c$-axis transport properties of nominally clean single-crystal cuprates than models where momentum is strictly conserved \cite{Ginsberg1994, Sheehy_2004}. Because random impurities break all spatial symmetries of the system, the form of the interlayer coupling is required to respect the crystal symmetry constraints only on average. One may thus expect that incoherent tunneling models will evade the difficulties noted above and produce a fully gapped topological phase near $\theta=45^\circ$. 

Based on a perturbative diagrammatic treatment within a simplified continuum model we show that the incoherent tunneling model indeed delivers the same phenomenology as the coherent model of Ref.\,\cite{Can2021} while respecting with the point group symmetries of the physical system. Importantly, we show that for sufficiently slowly varying disorder the ground state near $\theta=45^\circ$ is gapped in the $\mathcal{T}$-breaking region and topologically non-trivial. These results are then confirmed in a more realistic setting through a full numerical diagonalization of a lattice model with parameters chosen to reproduce the actual cuprate band structure in the vicinity of the Fermi level. The effect of interface inhomogeneity on Josephson effects in twisted bilayers was recently considered in Ref.\,\cite{volkov2021} where it was found that sufficiently strong disorder can leave the system in a topologically trivial state around 45$^\circ$. This is consistent with our deductions.

The paper is structured as follows. After summarizing the origin of the $\mathcal{T}$-broken phase in the language of group theory (Sec.~\ref{sec:group theory}), in Sec.~\ref{sec:model} we introduce a model of incoherent interlayer tunneling within a continuum framework and show that a fully-gapped  $\mathcal{T}$-broken phase emerges in the vicinity of $\theta=45^\circ$. The phase is topological as evidenced by chiral edge modes traversing the gap that appear in the disorder-averaged boundary spectral function. In Sec.~\ref{sec:latticemodel}, we supplement our continuum results with a lattice model that simultaneously captures the characteristic cuprate Fermi surface geometry, the moir\'e effects and disorder in interlayer coupling. Concluding remarks appear in Sec.~\ref{sec:conclusions}.


\section{Group theoretical discussion of $\mathcal{T}$-breaking in twisted cuprates}
\label{sec:group theory}

The phenomenology of $\mathcal{T}$-breaking in twisted cuprates can be captured by a two-component Landau-Ginzburg theory with complex order parameters $\Psi_1$ and  $\Psi_2$ with a relative phase $\varphi$.  The $\varphi$-dependent part of free energy is of the general form 
\begin{align}
    \mathcal{F}(\varphi) = -B \,  |\Psi_1\Psi_2| \cos \varphi + C \, |\Psi_1 \Psi_2|^2 \cos 2\varphi.
    \label{eq:LG}
\end{align}
Time reversal symmetry will be spontaneously broken whenever the free energy develops two minima that are related by $\mathcal{T}: \varphi \rightarrow -\varphi$. The Josephson coupling term, proportional to $\cos \varphi$, has only a single minimum at $\varphi=0$ or $\pi$, depending on the sign of $B$.
Presence the fourth order term proportional to $C\cos 2\varphi$ is therefore necessary to break $\mathcal{T}$. Additionally, one must have $C>0$, since otherwise the two minima of $\cos 2\varphi$ occur at $\varphi=0,\pi$ which map to themselves under $\mathcal{T}$. In Ref.\,\cite{Can2021} it was argued that $C$ is indeed positive based on microscopic mean-field calculations. We confirm that $C$ remains positive in the case of incoherent interlayer coupling in Sec.~\ref{sec:freeenergy}.

Given $C>0$, the fourth order term is minimized at $\varphi=\pm\pi/2$. Then, $\mathcal{T}$-breaking occurs as a consequence of the competition among the two terms in Eq.~\eqref{eq:LG}. Specifically, $\mathcal{T}$ will be broken when 
\begin{align}
    4 C \left| \Psi_1 \Psi_2 \right| > \left| B\right| \,.
    \label{eq:trsbreaking}
\end{align}
A special situation clearly arises if symmetry requires $B$ to vanish; then Eq.~\eqref{eq:trsbreaking} is guaranteed to be satisfied for any $C>0$. 

Next we describe a set of symmetry requirements under which the coefficient $B$ vanishes and the system is forced into the $\mathcal{T}$-broken phase. The order parameters $\Psi_1,\Psi_2$ transform according to irreducible representations (irreps) of the point group of the crystal. Two cases must be distinguished: (a) $\Psi_1$ and $\Psi_2$ transform under two \textit{different} 1D irreps or (b) $(\Psi_1,\Psi_2)$ transform under a 2D irrep \cite{Annett1990,Poniatowski2022}. The latter case is considered a generic pathway to $\mathcal{T}$-breaking that occurs immediately upon entering the SC phase. The former case generically yields two successive phase transitions with distinct critical temperatures, $T_{c}$ and $T_c'$, with $\mathcal{T}$-breaking setting at the lower one $T_c'$. Note that $T_c'$ can be zero or negative, in which case the $\mathcal{T}$-broken phase is physically not accessible \cite{Annett1990, Kaba.2019}.

The point symmetries of an untwisted cuprate bilayer form the point group $D_{4h}$. Here, the $d_{x^2-y^2}$ and $d_{xy}$ order parameters transform according to the 1D irreps $B_{1g}$ and $B_{2g}$, respectively. At arbitrary twist angle, inversion and mirror symmetries are broken and the point group reduces to $D_4$ with $d$-wave irreps $B_1$ and $B_2$. Thus, given the $d$-wave nature of the order parameter in cuprates, only pathway (b) to $\mathcal{T}$-breaking is possible and no definite symmetry-based arguments can be made.

Precisely at $\pf$, however, the symmetry group is enlarged to the non-crystallographic point group $D_{4d}$ which contains an additional $8$-fold improper rotation $S_8$ of the quasicrystalline lattice. Most notably, among the irreps of $D_{4d}$ \textit{only} the 2D $E_2$-irrep supports $d$-wave order. Thus, the Josephson coupling term $ -B \,  |\Psi_1\Psi_2| \cos \varphi$ must necessarily be absent at $45^\circ$. This is so because it descends from the $-B(\Psi_1\Psi_2^\ast +{\rm c.c.})$ term in the free energy which is not invariant under $S_8: (\Psi_1,\Psi_2)\rightarrow (\Psi_2,-\Psi_1)$. Thus, $\mathcal{T}$-breaking can be viewed as a fundamental consequence of the point group at $\theta= 45^\circ$. We summarize our key arguments as follows: Two-component order parameters that transform under a 2D irrep naturally break $\mathcal{T}$. At $45^\circ$ twist angle, because the point group of the bilayer is $D_{4d}$, any $d$-wave order parameter must necessarily transform under a 2D irrep. Therefore, the superconducting state breaks $\mathcal{T}$ right below $T_c$. 

The phase diagram of twisted bilayer cuprates derived in Ref.~\cite{Can2021} can then be understood from continuity arguments. It is expected that the $\mathcal{T}$-breaking phase will not be limited to the exact $45^\circ$-twist, but will extend to a range of twist angles in its vicinity. Since at twists slightly away from $45^\circ$ the order parameters transform under two 1D-irreps, two distinct critical temperatures are permitted and $\mathcal{T}$-breaking will no longer coincide with the critical temperature $T_c$ of the spontaneous $U(1)$-symmetry breaking.  This naturally leads to the wedge-shaped $\mathcal{T}$-broken domain in the phase diagram explicitly computed in Ref.~\cite{Tummuru2022}. Our symmetry arguments will be manifest in a microscopic description of the bilayer system.


\section{Incoherent tunneling}
\label{sec:model}

\subsection{Background and model definition}

Experimental measurements of the $c$-axis transport in bulk crystals of BSCCO and other cuprates, summarized for example in Ref.\,\cite{Ginsberg1994}, have been interpreted as evidence of interlayer tunneling dominated by disorder-mediated, incoherent processes. The $c$-axis superfluid stiffness, accessible through the measurement of the $c$-axis London penetration depth \cite{Hosseini1998, Panagopoulos2003}, provides a particularly clear evidence. Experimentally, the temperature dependence of the $c$-axis superfluid stiffness in clean single crystals was observed to follow an approximate power-law behavior $\rho_c = a-b T^\alpha$ with $\alpha\simeq2$ at low temperatures, whereas the in-plane stiffness showed a $T$-linear dependence \cite{Hardy1993}. The latter is the canonical behavior expected of a clean $d$-wave superconductor, reflecting the presence of low-energy excitations with a Dirac spectrum \cite{Hirschfeld1993}. Models with coherent tunneling between CuO$_2$  predict the same linear $T$-dependence for the $c$-axis stiffness \cite{Klemm1995}, in clear disagreement with experimental data. If the interlayer tunneling were dominated by the oxygen-assisted processes (the $g_1$ term in Eq.\ \eqref{gk}) then theory predicts $\rho_c = a-b T^5$ \cite{Xiang_1996}, again at variance with experiment. 
 
As demonstrated in Refs.\ \cite{Radtke1995,Radtke1996,Radtke_1997,Sheehy_2004} a description that captures the correct $\sim T^2$ scaling along the $c$-axis (while preserving the $T$-linear behavior in the $ab$ plane) can be given using the incoherent $c$-axis tunneling approach. In the following we shall review the relevant model and then apply it to the problem of a twisted bilayer.

A minimal model of the uncoupled bilayer system consists of the second-quantized Hamiltonian
\begin{align}
	\mathcal{H}_0 = \sum_{\mathbf{k}l}\Psi_{\mathbf{k}l}^\dagger
	H_{\mathbf{k}l}
	\Psi_{\mathbf{k}l}
\end{align}
where $l=1,2$ denotes the layer index, Nambu-Gorkov spinors $\Psi_{\mathbf{k}l}=(c_{\mathbf{k}l\uparrow },\, c_{-\mathbf{k}l\downarrow })^T$. In the BCS approximation, we have
\begin{align}
	H_{\mathbf{k}l}
	=
	\xi_{\mathbf{k}} \sigma_z 
	+ \Delta_{\mathbf{k}l}' \sigma_x
	- \Delta_{\mathbf{k}l}'' \sigma_y,
\end{align}
with Pauli matrices $\sigma_j$ acting in the Nambu space and $\Delta_{\mathbf{k}l}'$, $\Delta_{\mathbf{k}l}''$ denoting real and imaginary parts of the superconducting gap function. We adopt units such that $\hbar=e=k_B=m_e=a_0=1$, where mass is measured in units of electron mass $m_e$ and length scales in units of lattice constant $a_0$. To make the model tractable we assume a simple parabolic band dispersion given by $\xi_{\mathbf{k}}=\mathbf{k}^2/2m-\mu$ in each layer. (In Sec.\ \ref{sec:latticemodel} we consider a more realistic band structure and show that it leads to similar results.) The two superconducting $d$-wave order parameters are
\begin{align}
	\Delta_{\mathbf{k}1} &= \Delta e^{i\varphi/2} \cos(2\alpha_\mathbf{k} -
	\theta) 
	\nonumber
	\\
	\Delta_{\mathbf{k}2} &= \Delta e^{-i\varphi/2} \cos(2\alpha_\mathbf{k} +
	\theta)
	\label{eq:op}
	\,,
\end{align}
where $\varphi$ is the phase difference and $\alpha_\mathbf{k}$ denotes the polar angle of $\mathbf{k}$. 

The layers are coupled by the term
\begin{align}\label{tun}
	\mathcal{H}' = \sum_{\mathbf{kq}} \gamma_{\mathbf{q}} c_{\mathbf{k},1}^\dagger
	c_{\mathbf{k-q},2} + \textrm{h.c.}
\end{align}
and $\mathcal{H}=\mathcal{H}_0 + \mathcal{H}'$ constitutes the full model. The lack of momentum conservation in Eq.\ \eqref{tun} is the defining feature of the incoherent tunneling models and originates, physically, from the disorder present in the spacer layers separating the copper-oxygen planes. The disorder is captured via a set of Gaussian-distributed random variables $\gamma_\mathbf{q}$ of average $\overline{\gamma_\mathbf{q}}=0$ and variance given by
\begin{align}
	\overline{\gamma_\mathbf{q}^* \gamma_\mathbf{q+p}} &=
	\frac{1}{N}\frac{4\pi g^2}{3\Lambda^2}
	\delta_{\mathbf{p},0}e^{-\mathbf{q}^2/\Lambda^2} \,.
    \label{eq:incotunnterm}
\end{align}
The scale $\Lambda$ defines the characteristic momentum change that an electron undergoes when tunneling between the two layers. The factor $1/3$ is chosen to reproduce the phase diagram of the coherent model of Ref.~\cite{Can2021} in the limit $\Lambda\rightarrow 0$ for the same value of $g$. For simplicity we have neglected any $\theta$-dependence of the interlayer coupling although we expect the randomness to be stronger in twisted samples due to the increase in interface roughness, added strain, and moir\'e lattice modulations. 

The above form of incoherent interlayer tunneling is consistent with all lattice symmetries because $\gamma_\mathbf{q}$ vanishes on average. This constitutes the key difference to a coherent coupling of the form $\sum_{\mathbf{k}}(g\, c_{\mathbf{k},1}^\dagger c_{\mathbf{k},2} + \textrm{h.c.})$. As was pointed out in Ref.~\cite{Song2021}, at $45^\circ$ twist the two participating Cu $d$-orbitals transform under a 2D representation of $D_{4d}$ and a coherent tunneling term is therefore not invariant under $S_8: (c_{\mathbf{k},1},c_{\mathbf{k},2})\rightarrow (c_{\mathbf{k},2}, -c_{\mathbf{k},1})$. It thus vanishes by virtue of the same argument as the Josephson coupling in Eq.~\eqref{eq:trsbreaking}. This is indeed owed to the coincidence of the atomic Cu orbitals transforming under the same representation as the superconducting order parameters.

It is instructive to consider the limit $\Lambda \rightarrow 0$ of the incoherent tunneling in Eq.~\ref{eq:incotunnterm}. Here, one has $\overline{\gamma_\mathbf{q}^* \gamma_\mathbf{q+p}} = g^2\delta_{\mathbf{q},0}\delta_{\mathbf{p},0}/3$ and momentum is conserved in the interlayer tunneling process. Yet, $\Lambda \rightarrow 0$ is not the clean limit in the sense that, in real space, it corresponds to the case where macroscopic regions are correlated with the same random value of interlayer tunneling $g/\sqrt{3}$. From the viewpoint of disorder-induced incoherence, the random values of $g$ should only be correlated in the vicinity of an impurity which sets the appropriate scale for $1/\Lambda$. In the discussion below, the $\Lambda\rightarrow 0$ thus serves as an abstract but convenient reference point that connects the present model to the calculations in the original work \cite{Can2021}. We will refer to it as the \textit{coherent} limit.


\subsection{Free energy and phase diagram}
\label{sec:freeenergy}

\begin{figure}[t]
	\centering
	\includegraphics[width=\columnwidth]{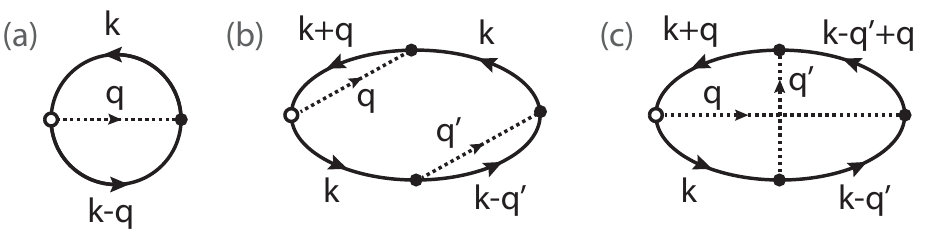}
	\caption{Diagrammatic expansion of the interlayer current (a) at order $g^2$ and (b-c) at order $g^4$. Full lines correspond to electronic propagators $G$ and dashed lines correspond to impurity vertices paired by disorder average. The open circle denotes the current vertex $j_\mathbf{q}$ defined in the main text and impurity vertices $\gamma_\mathbf{q}$ are given by black dots.}
	\label{fig:diagrams}
\end{figure}

The physics of $\mathcal{T}$-breaking is captured by the $\varphi$-dependence of the free energy. To determine the free energy we begin by implementing the global gauge transformation 
$
\left( c_{\mathbf{k}1},\,c_{\mathbf{k}2} \right)
\rightarrow
\left( c_{\mathbf{k}1}e^{i\varphi/4},\,c_{\mathbf{k}2}e^{-i\varphi/4} \right)
$
which moves the superconducting phase difference from the order parameters in Eq.~\eqref{eq:op} to $\mathcal{H}'$ according to
\begin{align}
	\mathcal{H}' \rightarrow 
	\mathcal{H}' =
	\sum_{\mathbf{kq}} \gamma_{\mathbf{q}} e^{i\varphi/2}c_{\mathbf{k},1}^\dagger
	c_{\mathbf{k-q},2} + \textrm{h.c.} \,.
\end{align}
In this gauge, the disorder-averaged interlayer current is given by
\begin{align}
	J &= \sum_{\mathbf{kq}} i e^{i\varphi/2}\overline{\gamma_\mathbf{q}
	\langle c_{\mathbf{k},1}^\dagger c_{\mathbf{k-q},2} \rangle} + \textrm{h.c.} 
	\nonumber
	\\
	&= \text{Tr}
	\left[ \overline{
		j_\mathbf{q} G(\mathbf{k},\mathbf{k-q},\omega_n)}
	\right]\,,
	\label{eq:current}
\end{align}
where $G(\mathbf{k},\mathbf{k'},\tau)=\langle T_\tau c_{\mathbf{k}}(\tau) c_{\mathbf{k'}}^\dagger(0) \rangle$ is the full imaginary time ordered Green's function of the disordered system and the current vertex is
\begin{align}
	j_\mathbf{q} =i\gamma_\mathbf{q} 
	\begin{pmatrix}
		0 & e^{i\sigma_z\varphi /2}\\
		-e^{-i\sigma_z\varphi /2} & 0
	\end{pmatrix} \,.
\end{align}
Note that the trace is to be performed over all momenta $\mathbf{k},\mathbf{q}$ and Matsubara frequencies $\omega_n=(2n+1)\pi/\beta$, in addition to interlayer and Nambu indices.

From the Josephson relation $J(\varphi)=2\partial \mathcal{F}(\varphi)/\partial \varphi$ one then obtains the functional dependence of the free energy on the interlayer phase difference $\varphi$ by simple integration. We expand Eq.~\eqref{eq:current} up to fourth order in $g$ while treating $\mathcal{H}'$ as a perturbation. Three different terms arise, which are diagrammatically represented in Fig.~\ref{fig:diagrams}. Panel (a) corresponds to the term 
\begin{align}
    J_c^{(2)} = \text{Tr}\left[\overline{j_{\mathbf{q}} G_0(\mathbf{k-q},\omega_n)
    	u_{-\mathbf{q}} G_0(\mathbf{k},\omega_n)}
    \right] \,
    \label{eq:j1}
\end{align}
which is quadratic in $g$. Here, $G_0(\mathbf{k},\omega_n)=\left( i\omega_n - H_{\mathbf{k}} \right)^{-1}$ is the unperturbed, translationally invariant Green's function with $H_{\mathbf{k}}={\rm diag}(H_{\mathbf{k}1},H_{\mathbf{k}2})$ and
\begin{align}
	u_\mathbf{q} =\gamma_\mathbf{q} 
	\begin{pmatrix}
		0 & \sigma_ze^{i\sigma_z\varphi /2}\\
		\sigma_z e^{-i\sigma_z\varphi /2} & 0
	\end{pmatrix} \,.
\end{align}
is the impurity vertex. The disorder average acts on $\gamma_\mathbf{q}$ factors and is performed according to Eq.\ \eqref{eq:incotunnterm}.
The diagrams in Fig.~\ref{fig:diagrams}(b-c) represent terms of order $g^{4}$:
\begin{equation}
\begin{aligned}
	J_c^{(4)} &= 
	2\, \text{Tr} \left[ j_{\mathbf{q}} G_0(\mathbf{k},\omega_n)
		u_{\mathbf{q}'} G_0(\mathbf{k-q'},\omega_n)
		u_{-\mathbf{q}'}
		\right.
		\nonumber
		\\
		& \quad\quad
		\times
		\left.
		G_0(\mathbf{k},\omega_n)
		u_{-\mathbf{q}}G_0(\mathbf{k+q},\omega_n)
	\right] \\
	&\quad +
	\text{Tr}\left[ 
	    j_{\mathbf{q}} 
	    G_0(\mathbf{k},\omega_n)
		u_{\mathbf{q}'} 
		G_0(\mathbf{k-q'},\omega_n)
		u_{-\mathbf{q}}
		\right.
		\nonumber
		\\
		& \quad\quad
		\times
		\left.
		G_0(\mathbf{k-q'+q},\omega_n)
		u_{-\mathbf{q}'}G_0(\mathbf{k+q},\omega_n)
	\right]
\end{aligned}
\end{equation}
where impurity averaging is assumed but not explicitly shown for clarity of notation. Evaluating the traces, we obtain the current of the form 
\begin{align}
	J = J_c^{(2)} + J_c^{(4)} = J_{c1}(\theta) \sin \varphi - J_{c2}(\theta) \sin 2\varphi \,,
	\label{eq:jc}
\end{align}
with coefficients, to lowest order of $g$,
\begin{align}
    J_{c1}&=
	4
	\sum_{n\mathbf{k}} j_{n\mathbf{k}}
    \label{eq:jc1}
    \\
	J_{c2}&=
	8
	\sum_{n\mathbf{k}}
	j_{n\mathbf{k}}^2
	+
	4\sum_{n\mathbf{kqq'}} 
	\left|\gamma_\mathbf{q} \right|^2
        \left|\gamma_\mathbf{q'} \right|^2
        \label{eq:jc2}
        \\
        &
        \times
         f_{n\mathbf{k},1}f_{n\mathbf{k+q+q'},1}
        f_{n\mathbf{k+q},2}f_{n\mathbf{k+q'},2}
	\nonumber
\end{align}
Here, we have defined
\begin{align}
\label{eq:convolution}
    j_{n\mathbf{k}} &= \left(f_{n\mathbf{k},1} * \left|\gamma_{n\mathbf{k}} \right|^2 \right)
    f_{n\mathbf{k},2}
    \\
    f_{n\mathbf{k},l} &= \frac{\Delta_{\mathbf{k},i}}{\omega_n^2+E_{\mathbf{k},l}^2}
    \label{eqn:f-function}
\end{align}
and $(*)$ denotes a convolution integral, $a_{\mathbf{k}}*b_{\mathbf{k}} = \sum_{\mathbf{q}}a_{\mathbf{q}}b_{\mathbf{k-q}}$. The quasiparticle dispersion of the unperturbed bands is given by $E_{kl} = \sqrt{\xi_{\mathbf{k}}^2+\Delta_{\mathbf{k}l}^2}$.

From Eq.~\eqref{eq:jc} one obtains the free energy
\begin{align}
    2\mathcal{F}=-J_{c1}(\theta) \cos \varphi + \frac{J_{c2}(\theta)}{2} \cos 2\varphi +\textrm{const}\,.
\end{align}
The $\mathcal{T}$-breaking phase transition occurs as a consequence of competition between $\cos \varphi$ and $\cos 2 \varphi$ terms. Clearly, $J_{c2}>0$ and the ground state acquires a finite phase difference for 
\begin{align}
    2J_{c2}>|J_{c1}|,
    \label{eq:crit}
\end{align}
where it spontaneously breaks $\mathcal{T}$. From our discussion in Sec.~\ref{sec:group theory} it follows that that $J_{c1}$ must vanish at twist of $\theta=\pi/4$. Explicitly, one can see this result as follows. The functions $E_{\mathbf{k},l}, \, \left|\gamma_\mathbf{k} \right|^2$ transform under the $A_{1g}$ irrep of $D_{4h}$ whereas the $f_{\mathbf{k},1},f_{\mathbf{k},2}$ transform under $B_{1g}$ and $B_{2g}$, respectively. We note that convolution with the $A_{1g}$-symmetric impurity distribution $\left|\gamma_\mathbf{k} \right|^2$ does not change the symmetry of the convolution integral. Hence, $j_{\mathbf{k}}$ transforms under $B_{1g}\otimes B_{2g}=A_{2g}$ and all terms in Eq.~\eqref{eq:jc1} average to zero at $\pf$-twist. However, $j_{\mathbf{k}}^2$ is $A_{1g}$-symmetric and $J_{c2}$ will consequently be finite and positive. Thus it is clear that condition \eqref{eq:crit} is generally satisfied  at $\theta=45^\circ$ and $\mathcal{T}$ will always be broken as soon as the system enters the SC state below $T_c$.

\begin{figure}[t]
	\centering
	\includegraphics[width=\columnwidth]{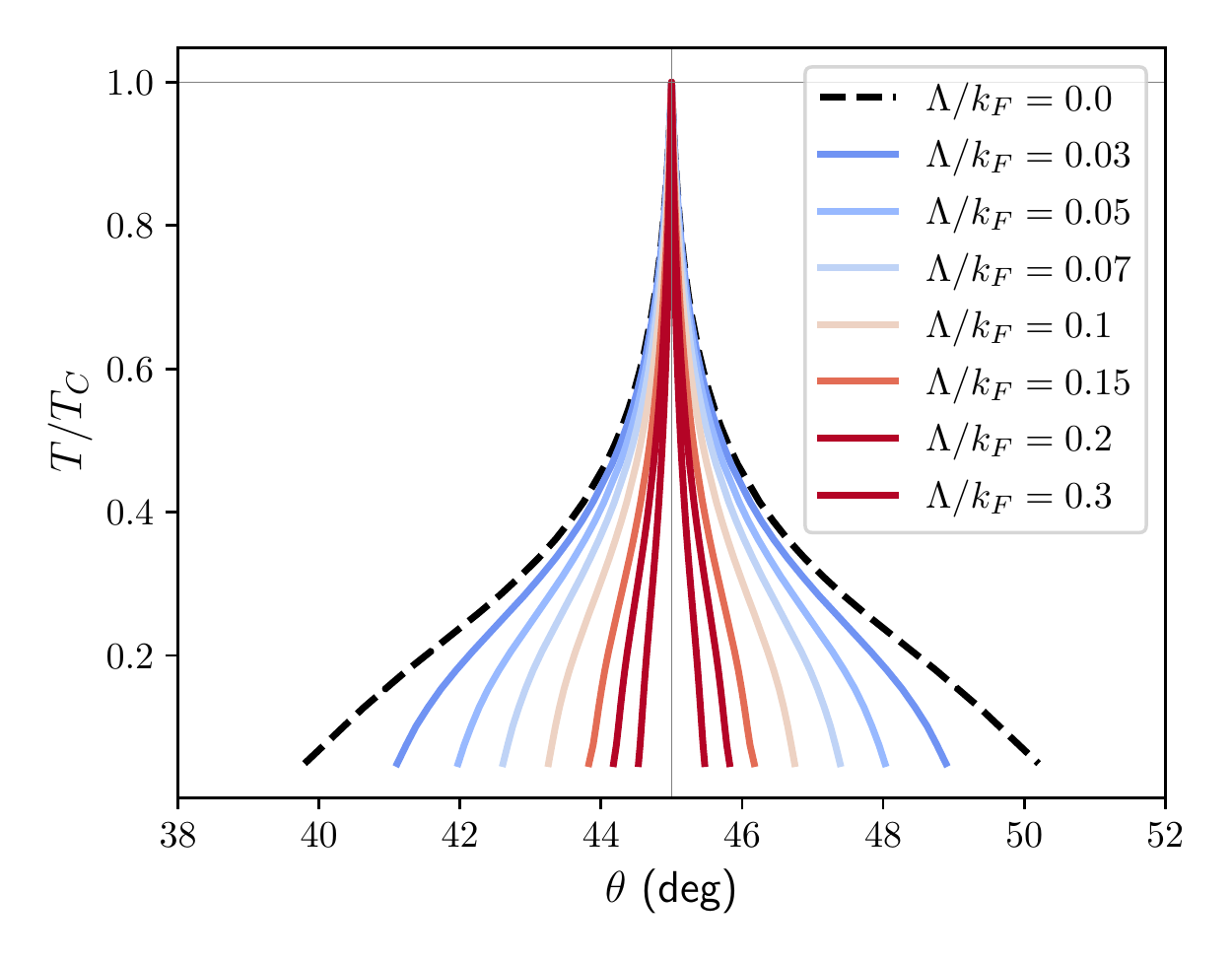}
	\caption{Phase diagram of incoherently coupled twisted bilayer cuprates. For a given $\Lambda$, the inside of the cone-shaped region breaks $\mathcal{T}$. Black-dashed lines mark the phase boundary in the clean limit, previously introduced in \cite{Can2021}. For increasing degree of momentum non-conservation $\Lambda$, the $\mathcal{T}$-breaking phase boundaries shrink towards $45^\circ$.}
	\label{fig:phasediagram}
\end{figure}

We conclude that impurity-mediated tunneling must not qualitatively change the $\mathcal{T}$-breaking phase diagram relative to the model of Ref.\ \cite{Can2021}. The incoherent tunneling, however, shifts the phase boundaries. As shown in Appendix \ref{apdx:lambda-scaling}, $J_{c1}\sim 1/\Lambda$ and $J_{c2}\sim 1/\Lambda^2$. Since $J_{c1}$ is only weakly dependent on $\theta$, and $J_{c1}$ vanishes linearly around $45^\circ$ twist, it follows from Eq.~\eqref{eq:crit} that the width of the $\mathcal{T}$-breaking phase space is proportional to $J_{c2}(\theta=0)/J_{c1}(\theta=0) \sim 1/\Lambda$. In the perfectly incoherent limit, $\Lambda\rightarrow\infty$, the free energy becomes independent of $\varphi$ and the $\mathcal{T}$-breaking phase disappears.

To quantitatively ascertain the effect of incoherent tunneling on the phase diagram, we numerically evaluate the coefficients $J_{ci}$. In principle, all Matsubara sums can be evaluated analytically, at the cost of removing the simple convolution structure in Eq.~(\ref{eq:convolution}). This leaves three remaining momentum integrals to be numerically evaluated at complexity $\mathcal{O}(N^3)$ where $N$ is the number of $\mathbf{k}$-points of the 2D mesh used to perform the integrals. A more efficient approach is to exploit the convolution structure of Eq.~(\ref{eq:convolution}) using the fast Fourier transform (fft) algorithm and numerically evaluate $M$ Matsubara frequencies, affording evaluation of diagrams Fig.~\ref{fig:diagrams}(a-b) at order $\mathcal{O}(MN\log N)$. The crossed diagram Fig.\ \ref{fig:diagrams}(c) does not possess a convolution structure. As we show in Appendix \ref{apdx:complexity}, it can be evaluated at a cost of $\mathcal{O}(MN^2)$.

The resulting phase diagram is shown in Fig.~\ref{fig:phasediagram} for coupling strength $g=\SI{10.5}{\milli\electronvolt}$ and several values of $\Lambda$. We see that the $\mathcal{T}$-breaking phase space is largest in the `clean' limit $\Lambda\rightarrow 0$ where it extends between $(45\pm 6)^\circ$ at $T=0$. Increasing $\Lambda$ gradually reduces the extent of the $\mathcal{T}$-broken phase which eventually vanishes in the perfectly incoherent limit when $\Lambda \sim k_F$, i.e. when impurity correlations are on the scale of the lattice constant. Physically, this occurs because at this level of incoherence the Cooper pair essentially looses all memory of its momentum structure in the process of tunneling between layers.


\subsection{Spectral gap and topological superconductivity}

Having discerned the fate of the $\mathcal{T}$-breaking phase in the presence of impurity-mediated tunneling we proceed to examine the topological properties of the resulting ground state. In the clean limit, $\mathcal{T}$-breaking establishes a topological phase with Chern number $\mathcal{C}=4$ \cite{Can2021}. Since the disordered model is connected to the clean case by taking the limit $\Lambda \rightarrow 0$, it is reasonable to expect the same $\mathcal{C}=4$ phase as long as the quasiparticle gap does not close.

Here, we show that these expectations are indeed met. To this end, we evaluate the Green's function
\begin{align}
    G(\mathbf{k},\omega_n) = [G_0-\Sigma(\mathbf{k},\omega_n)]^{-1}
    \label{eq:gfd}
\end{align}
in the Born approximation where
\begin{align}
    \Sigma_{\tau\tau'} &=
    \sum_{\mathbf{q}} u_{\mathbf{q}} \, G_0(\mathbf{k-q},\omega_n)\,  u_{\mathbf{-q}}
    \\
      &=-\delta_{\tau,\tau'}\,f_{\mathbf{k},\bar{\tau}} (i\omega_n + \xi_{\mathbf{k}} \sigma_z + e^{i\tau\sigma_z \varphi}\Delta_{\mathbf{k},\bar{\tau}} \sigma_x ) 
      * \left|\gamma_{\mathbf{k}}\right|^2
      \nonumber \,.
 \end{align}
 with layer-indices $\tau=\pm 1$. Here, we regularized the continuum model on a square lattice using 
 \begin{align}
     \xi_\mathbf{k}&=-2t (\cos k_x +\cos k_y) - \mu
     \nonumber
     \\
     \Delta_{\mathbf{k},\tau}&=  \Delta [(\cos k_x - \cos k_y) \cos \theta 
     \label{eq:alignedgap}
     \\
     & \qquad\quad
     + \,\tau \sin k_x \sin k_y \, e^{-i\varphi}\sin \theta ]
     \nonumber
 \end{align}
 with parameters chosen to match the continuum model.
 
Following the method introduced in Ref.~\cite{Pinon.2020}, we compute a spatially resolved Green function
\begin{align}
     G_B(x, k_y,\omega_n) &= G(x, k_y) - G(x, k_y) 
     T(k_y)
     G(-x, k_y) 
     \nonumber
     \\
     T(k_y) &= \left[ 
     \frac{1}{\sqrt{N}} \sum_{k_x} G(k_x, k_y)
     \right]^{-1} 
     \label{eq:transfm}
\end{align}
in the presence of a strong repulsive potential at $x=0$ which simulates an edge and thus allows us to inspect the edge modes of the disordered system. We outline the method and give a derivation of Eq.~\eqref{eq:transfm} in Appendix~\ref{apdx:surfacegf}. 

In Fig.~\ref{fig:edgemode} we plot the analytically continued boundary spectral function
\begin{align}
    A_B(x,k_y,\omega) = -\frac{1}{\pi} \text{Im} \left[G_B(x,k_y,-i\omega+\eta)\right]
\end{align}
at the edge $(x=1)$ as well as the bulk spectral function. We clearly observe two chiral edge modes traversing the bulk gap thus confirming the non-trivial topology of the system. The edge modes display a degeneracy in the layer degree of freedom, suggesting that the model is in a topological phase with Chern number $\mathcal{C}=4$. The bulk gap is reduced but remains finite as $\Lambda$ increases.

\begin{figure}[t]
	\centering
	\includegraphics[width=\columnwidth]{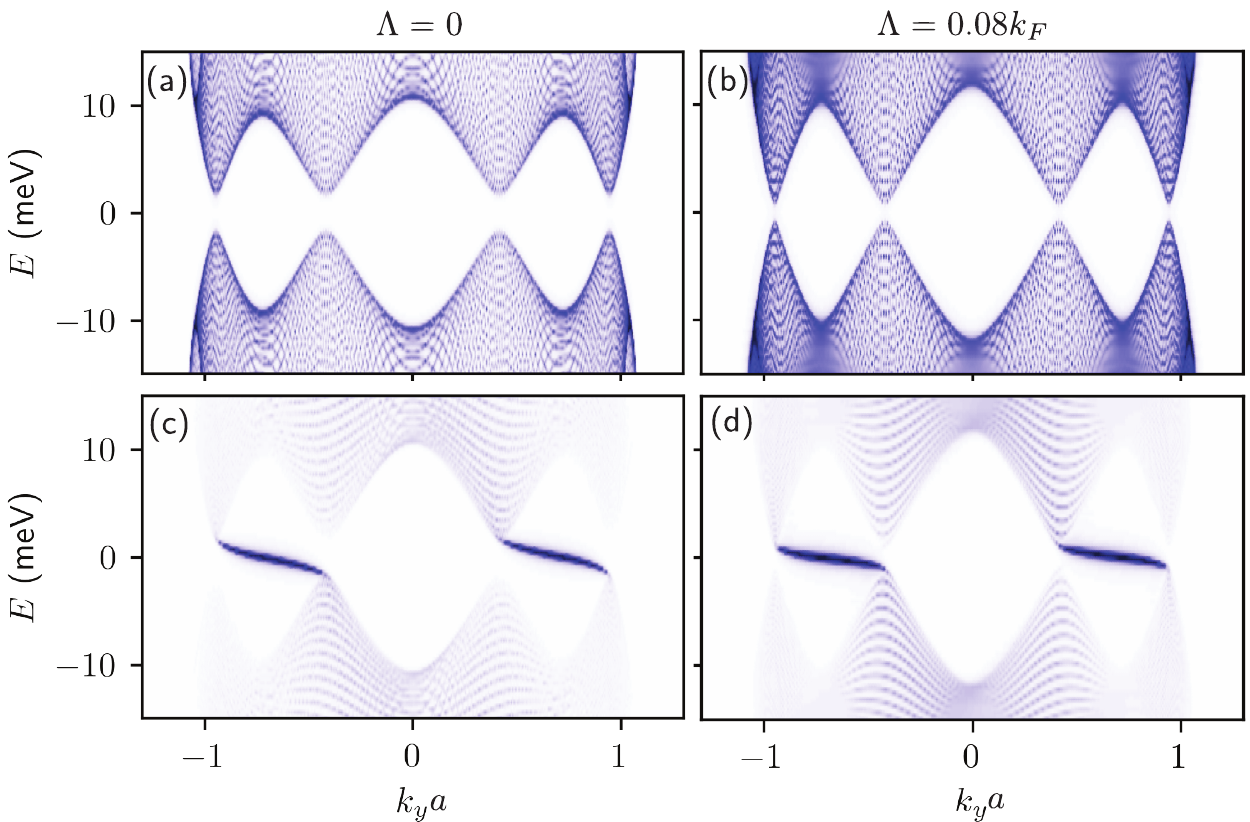}
	\caption{Bulk (a-b) and boundary (c-d) spectrum for incoherently coupled cuprate bilayers with $\Lambda=0$ (left) and $\Lambda/k_F=0.08$ (right) at $45^\circ$ twist angle. The spectrum shows chiral edge modes traversing the bulk gap which is reduced but finite for increased $\Lambda$. Edge modes, which are in fact degenerate, indicate a Chern number $\mathcal{C}=4$. }
	\label{fig:edgemode}
\end{figure}


\section{Lattice model}
\label{sec:latticemodel}

\begin{figure*}[t]
  \includegraphics[width=16cm]{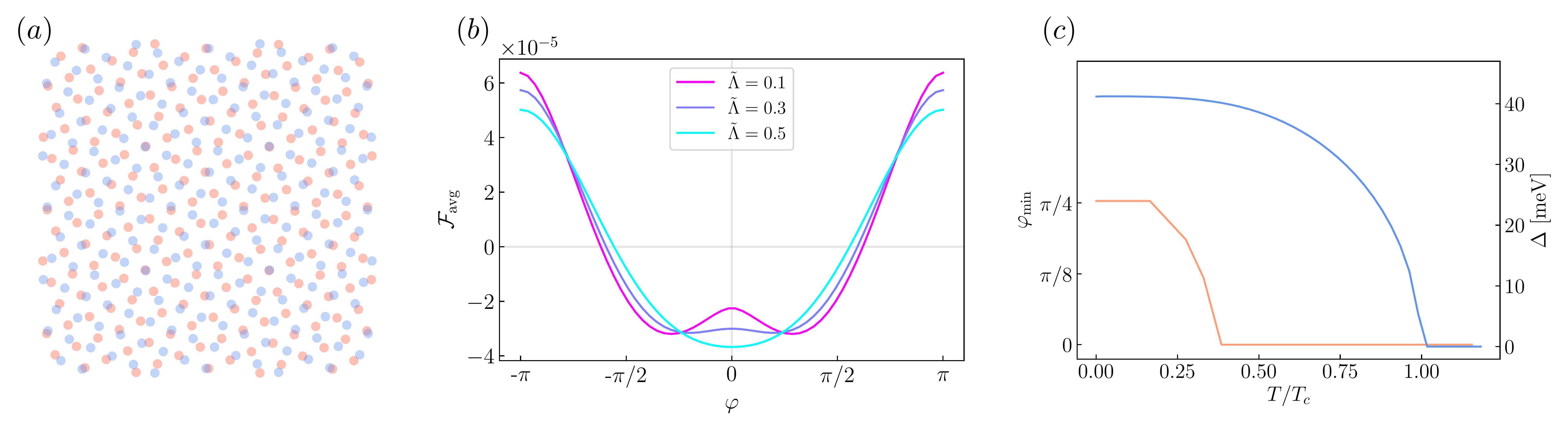}
  \caption{(a) Illustration of the geometry of the bilayer lattice model at an incommensurate angle of $\sim 43^\circ$. (b) Disorder averaged free energy of the bilayer at zero temperature as a function of the phase difference. The minima $\varphi_{\rm min}$ are situated away zero at small disorder strengths. (c) Dependence of the order parameter amplitude and phase as a function of temperature for $\tLambda=0.2$. One could view it as a vertical cut at a specific twist in the phase diagram of Fig.~\ref{fig:phasediagram}, with the onset of a non-zero phase marking the phase boundary.}
  \label{fig:latt_F}
\end{figure*}

So far we have looked at the role of disorder in a continuum formulation of a twisted bilayer. The two-site unit cell of the regularized model allowed for analytical expressions for the layer Green's function to which we have systematically added incoherent interlayer tunneling and calculated the free energy up to fourth order in $g$. Another approach to tackle the problem and corroborate the results in a more general setting is to perform a BCS mean field theory on two twisted square lattices that represent the two CuO planes. In this case one can incorporate more realistic band structures but it is also harder to obtain the Green's functions analytically using Feynman diagrams. The reason is twofold: For one, at an arbitrary twist the lattice model is not commensurate. Secondly, at commensurate twist angles close to $45^\circ$, the moir\'{e} unit cell contains many sites. To get around this, we perform a brute force disorder average wherein several disorder realizations with the same microscopic parameters are taken into account. While it limits us to real space, such a treatment is exact because all orders in perturbation theory are implicitly accounted for.

For each layer, we consider a square lattice Hubbard model with nearest neighbor density-density interactions such that a mean-field decoupling produces a $d$-wave order parameter. Including the interlayer tunneling processes with amplitudes $g_{ij}$, the bilayer is described by
\begin{eqnarray}
    &\cH  = & -t \sum_{\langle ij \rangle \sigma l} c^\dag_{i \sigma l} c_{j \sigma l} - t'\sum_{\langle \langle ij \rangle \rangle \sigma l} c^\dag_{i \sigma l} c_{j \sigma l} 
    - \mu \sum_{i \sigma l} n_{i \sigma l} \nonumber \\
    &+&\sum_{\langle ij \rangle l}\left(\Delta_{ij, l}c^\dag_{i\uparrow
      l}c^\dag_{j\downarrow l}+{\rm h.c.}\right) 
    - \sum_{i j \sigma} g_{ij} c^\dag_{i \sigma 1} c_{j \sigma 2},
    \label{eq:hm_latt}
\end{eqnarray}
where $l$ is a layer index, $t$ ($t'$) is the (next-)nearest-neighbor hopping amplitude, $\mu$ is the chemical potential that controls on-site particle density $n_{i \sigma l}$ and $\Delta_{ij,l}$ denotes the complex order parameter on the bond connecting sites $i$ and $j$ on layer $l$. Considering a fully coherent interlayer tunneling, Ref.~\cite{Can2021} employs a circularly symmetric, exponentially decaying form $g_{ij} = e^{-(r_{ij}-c)/\rho}$ which connects sites $i$ and $j$ separated by $\br_{ij}$. Therein $c$ in the interlayer separation and $\rho$ is defined by the radial extent of the participating orbitals. The twist angle $\theta$ between the layers determines connectivity and the strength of the interlayer tunnelings. The free energy of this model shows a double-well structure for twist angles around $45^\circ$ \cite{Can2021}. 

To incorporate incoherent processes, we introduce a random tunneling factor that vanishes on average, but encodes the correlation between different processes depending on spatial separation. That is, 
\begin{equation}
    g_{ij} = g_{\bR} ~ e^{-(r_{ij}-c)/\rho}    
    \label{eq:g_inco}
\end{equation}
where $\bR = (\br_{i} + \br_{j})/2$ denotes the center of mass location of the hopping and 
\begin{align}
    \overline{g_{\bR}} &= 0, \nonumber \\
    \overline{g_{\bR}g_{\bR'}} &= g^2 \exp\left[-\frac{\tLambda^2}{4} (\bR-\bR')^2 \right].
\end{align}
Analogous to the parameter $\Lambda$ in the continuum model, $\tLambda$ sets the length scale for the correlation between different tunneling amplitudes and is indicative of disorder strength. We distinguish the two simply because of the slightly differing definitions. To simulate the Fermi surface of optimally doped BSCCO with hole pocket around $(\pi,\pi)$, we set $t=153$meV, $t' = -0.45t$ and $\mu = -1.35t$ \cite{Bille2001}. Further, we choose $c=2.2$ and $\rho=0.4$ (in units of the lattice constant) to set interlayer distances. The $d$-wave order parameters in cuprates originates in the CuO planes and the interlayer coupling is a minor perturbation that does not influence the order parameter magnitude. In other words, temperature dependence of the gap in each layer is independent of twist and coupling strength strength $g$, which we peg at 20meV. Therefore, we use a $\Delta$ calculated self-consistently in a monolayer, which has a maximum of $\sim 40$meV at 0K in accordance with experimental findings in cuprates \cite{Dama2003, Fischer2007}.

To look for $\cT$-breaking we examine the free energy of the system, which can be calculated from the eigenvalues $E_i$ of the BdG Hamiltonian \eqref{eq:hm_latt}:
\begin{equation}
    \cF_{\rm BdG}= \sum_i E_i - 2 k_B T \sum_{i}\ln\left[2\cosh{(E_{i}/2 k_B T)}\right].
\end{equation}
In particular, for a given twist $\theta$ and disorder parameter $\tilde\Lambda$, we draw from the distribution \eqref{eq:g_inco} and average the free energy over 50 independent realizations. We choose a square bilayer sample as shown in Fig.~\ref{fig:latt_F}(a), but the results are independent of the shape. Further, the exact number of sites in the system depends on the cut and the twist angle, but the free energy does not show an appreciable change beyond $\sim 900$ sites per layer. In agreement with the continuum model, Fig.~\ref{fig:latt_F}(b) shows that the presence of $\cT$-breaking free energy minima is controlled by $\tilde\Lambda$. Namely, small values of $\tilde\Lambda$ support the $\cT$-broken ground state while larger values do not. 


\section{Conclusions}
\label{sec:conclusions}

Twisted bilayers of high-$T_c$ cuprates hold the potential for realizing topological superconductivity, wherein a topological gap is spontaneously induced. As per the symmetry informed momentum space form factors, which determine the electron hopping between interlayer Cu atoms, the tunneling amplitude vanishes along the nodal directions and a spectral gap may not appear. In this work we highlight that an important aspect to consider in such an analysis is the disorder mediated tunneling. Not only does disorder appear naturally due to oxygen doping and interfacial defects, but incorporating momentum non-conservation has been shown to better represent experimental data in clean single crystals.

Using perturbative diagrammatic calculations and disorder averaging on the lattice, we find that an experimentally motivated incoherent tunneling model that respects all point group symmetries of the physical system gives rise to a qualitatively similar phase diagram as obtained in Ref.\,\cite{Can2021}. Specifically we find a substantial range of twist angles around $45^\circ$ and temperatures where spontaneous $\cT$-breaking occurs and produces a fully gapped topological phase with non-zero Chern number. The angular extent of the $\cT$-broken phase depends on the disorder length scale $\Lambda^{-1}$ where the coherent limit $\Lambda\to 0$ recovers the phase diagram of Ref.\ \cite{Can2021} and increasing $\Lambda$ corresponds to a shrinking extent of the topological phase. Only when the incoherence length scale is comparable to the Fermi momentum, the twist angle for spontaneous $\cT$ breaking is reduced to exactly $\pf$. 

From an experimental point of view, the inhomogeneity due to oxygen doping the BiO planes of untwisted BSCCO was found to be correlated over $\approx 14 \si{\angstrom}$ \cite{Pan2001}. Since the CuO plane lattice constant is $\approx 5 \si{\angstrom}$, that amounts to a correlations over 3 unit cells, i.e., $\tLambda \approx 0.3$. In a twisted geometry, one may expect the characteristic length scale to decrease and, hence, the estimate for $\tLambda$ could shift up. That said, the role of complex atomic arrangements, moir\'e length scales and strong correlations are difficult to incorporate into such a heuristic reasoning. One would probably have to await data from complementary experimental probes, such as transport and optical response, to discern the nature of the superconducting state around $\pf$.

\begin{figure}[t]
	\centering
	\includegraphics[width=\columnwidth]{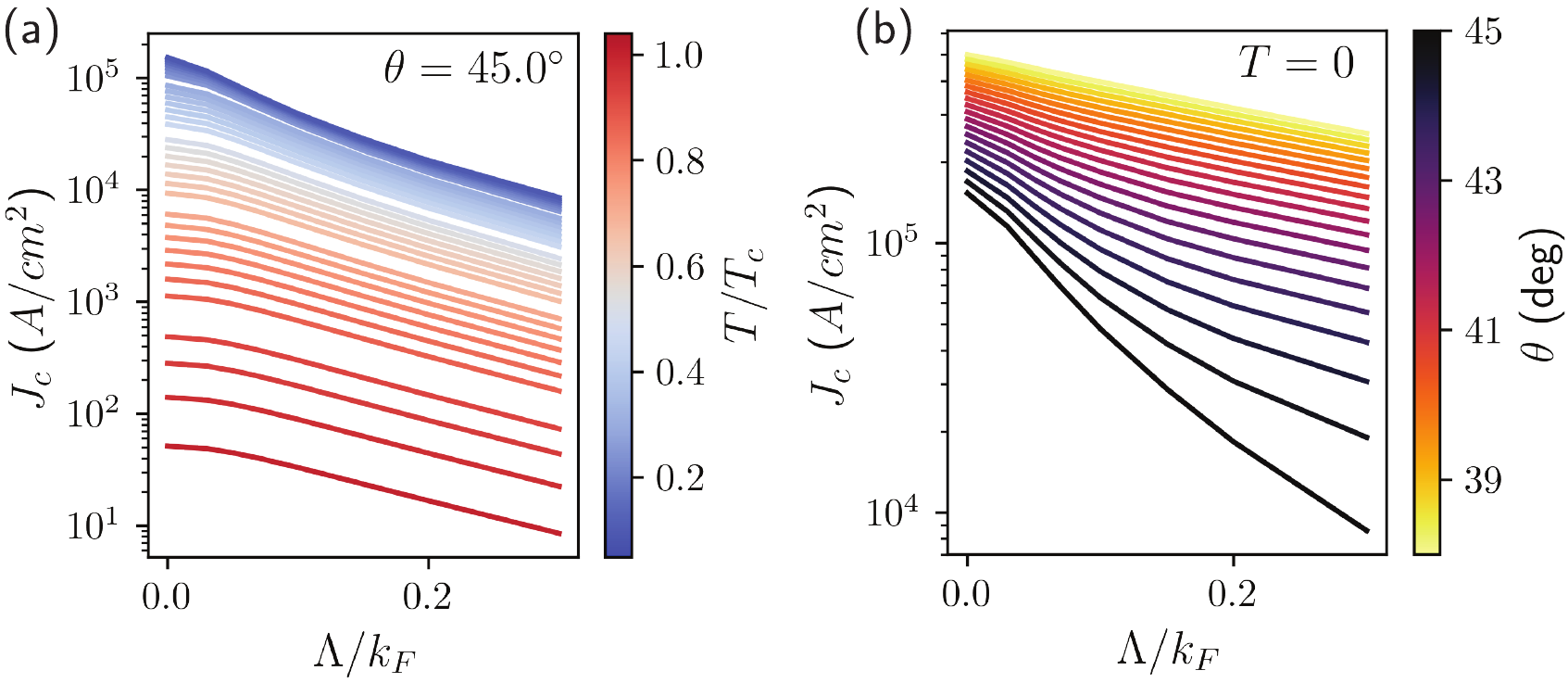}
	\caption{Critical current $J_c$ of the twisted bilayer as a function as interlayer coherence scale $\Lambda$. Incoherence significantly reduces $J_c$. The color scale denotes temperature $T$ in panel (a) and twist angle $\theta$ in (b).}
	\label{fig:jc}
\end{figure}

It was noted in Ref.\,\cite{Song2021} that the measured critical current density $J_c$ in both twisted and untwisted Bi2212 is about factor of 500 smaller than the theory prediction based on the slave-boson mean field theory of a $t$-$J$ model used in that study. We checked that a similar discrepancy occurs in the calculation using BCS mean field theory of Ref.\,\cite{Can2021}. As indicated in Fig.\,\ref{fig:jc} the discrepancy is somewhat reduced in the incoherent tunneling model (by about one order of magnitude at large $\Lambda$) but nevertheless significant disagreement with experiment persists. As noted in Ref.\,\cite{Sheehy_2004} this is a known problem that affects superconductors in the cuprate family and becomes increasingly severe in the underdoped part of their phase diagram. A phenomenological fix can be implemented \cite{Sheehy_2004} by restricting the momentum sums in the expression for $J_c$ to patches of linear size $\sim x$ (the hole doping) around the nodal points of the $d$-wave order parameter. This modification leaves the temperature dependence of $\rho_{ab}(T)$ and $\rho_{c}(T)$ unchanged, but reduces their $T=0$ magnitude to experimentally observed values. It similarly fixes the problem with $J_c$. As with many aspects of cuprates a truly microscopic understanding of this phenomenon remains a challenge to the theory community. With regards to twisted cuprate bilayers it would be interesting to explore the effect of the phenomenological fix outlined above on the phase diagram but we leave this to future work. 


\section*{Acknowledgments}

We are grateful to S. Egan, O. Can, X. Cui, \'E. Lantagne-Hurtubise, X.-Y. Song, A. Vishwanath, P. Volkov and S.Y.F. Zhao for helpful discussions and correspondence. This research was supported in part by NSERC and the Canada First Research Excellence Fund, Quantum Materials and Future Technologies Program. We thank the Max Planck-UBC-UTokyo Center for Quantum Materials for fruitful collaborations and financial support. R.H. acknowledges the Joint-PhD program of the University of British Columbia and the University of Stuttgart.


\bibliography{ref}


\appendix

\section{$\Lambda$-dependence of diagrams}
\label{apdx:lambda-scaling}

\begin{figure}[t]
    \includegraphics[width=\columnwidth]{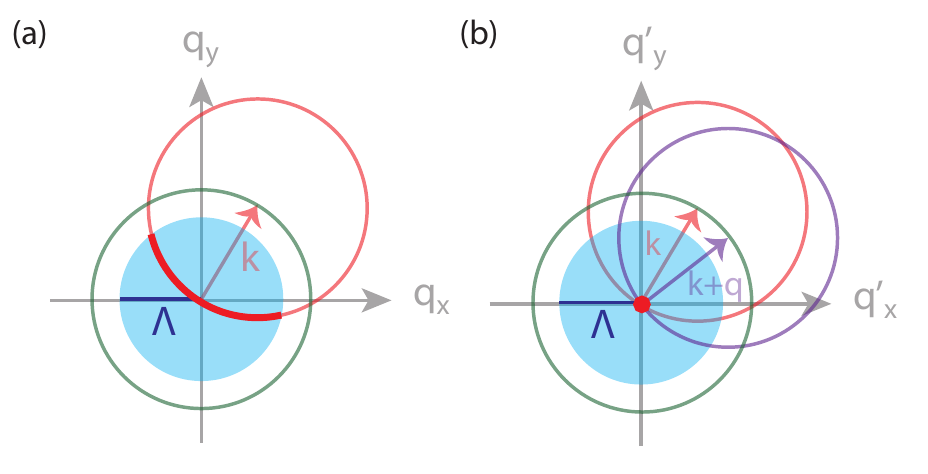}
    \caption{Graphical representation of the momentum summation constraints used to estimate $\Lambda$-dependence of the diagrams in Fig.~\ref{fig:diagrams}. In (a) the summation is restricted to the thick red segment of length $\Lambda$, whereas in (b) it is restricted to a small region around the origin, independent of $\Lambda$.}
    \label{fig:lambda}
\end{figure}

We can estimate the $\Lambda$-dependence of the three diagrams in Fig.~\ref{fig:diagrams}. First consider diagram Fig.~\ref{fig:diagrams}(a) which contains the coefficient
\begin{align}
    J_{c1}=4\sum_{n\mathbf{kq}}f_{n\mathbf{k},2}f_{n\mathbf{k+q},1}\left|\gamma_{\mathbf{q}}\right|^2 \,.
    \label{eqn:l1}
\end{align}
The function $f_{n\mathbf{k},i}$ is defined in Eq.~\eqref{eqn:f-function}. It is strongly peaked at the Fermi surface. Thus, in the above equation the term $f_{n\mathbf{k},2}$ restricts the summation to $|\mathbf{k}|\approx k_F$. The sum over $\mathbf{q}$ is constrained by the terms $f_{n\mathbf{k+q},1}$ and $\left|\gamma_{\mathbf{q}}\right|^2 $. For a given $\mathbf{k}$, the summation over $\mathbf{q}$ of the former term $f_{n\mathbf{k+q},1}$ may be visualized by a circle of radius $k_F$, displaced by $\mathbf{k}$, as illustrated in Fig.~\ref{fig:lambda}. The latter term corresponds to the blue shaded area $|\mathbf{q}|<\Lambda$. The combined constraints then restrict the summation to the red segment of length $2\Lambda$. Since $\left|\gamma_{\mathbf{q}}\right|^2 \sim 1/\Lambda^2$, the overall dependence on the momentum scattering scale is 
\begin{align}
    J_{c1}\sim \frac{1}{\Lambda}
\end{align}
The diagram in Fig.~\ref{fig:diagrams}(b) factors into two expressions of the type shown in Fig.~\ref{fig:diagrams}(a), as is clear from Eq.~\eqref{eq:jc2}. It is therefore proportional to $1/\Lambda^2$.

The crossed diagram Fig.~\ref{fig:diagrams}(c) is given by
\begin{align}
    J_{c2}^{\text{crossed}}&=
	4\sum_{n\mathbf{kqq'}} 
         f_{n\mathbf{k},1}
        f_{n\mathbf{k+q},2}
        f_{n\mathbf{k+q'},2}
        f_{n\mathbf{k+q+q'},1}
        \left|\gamma_\mathbf{q} 
       \gamma_\mathbf{q'} \right|^2.
        \nonumber      
\end{align}
The sum over $\mathbf{q}$ entails a factor of $\Lambda$ following the same argument as described above. The sum over $\mathbf{q}'$ is constrained by the terms $f_{n\mathbf{k+q'},2} f_{n\mathbf{k+q+q'},1}$, shown as the intersection of red and purple circles in Fig.~\ref{fig:lambda}(b). For large enough $\Lambda$, the intersection is not further constrained by the blue shaded region and the summation is independent of $\Lambda$. With $\left|\gamma_\mathbf{q} \gamma_\mathbf{q'} \right|^2\sim 1/\Lambda^4$, one then arrives at
\begin{align}
    J_{c2}^{\text{crossed}} \sim \frac{1}{\Lambda^3} \,.
\end{align}
These arguments break down for small enough $\Lambda$ where the width of the circular constraints $f_{n\mathbf{k},i}$ becomes relevant. Importantly, the diagrams Fig.~\ref{fig:diagrams}(b-c) become identical in the limit $\Lambda\rightarrow 0$.

\begin{figure}[t]
    \includegraphics[width=\columnwidth]{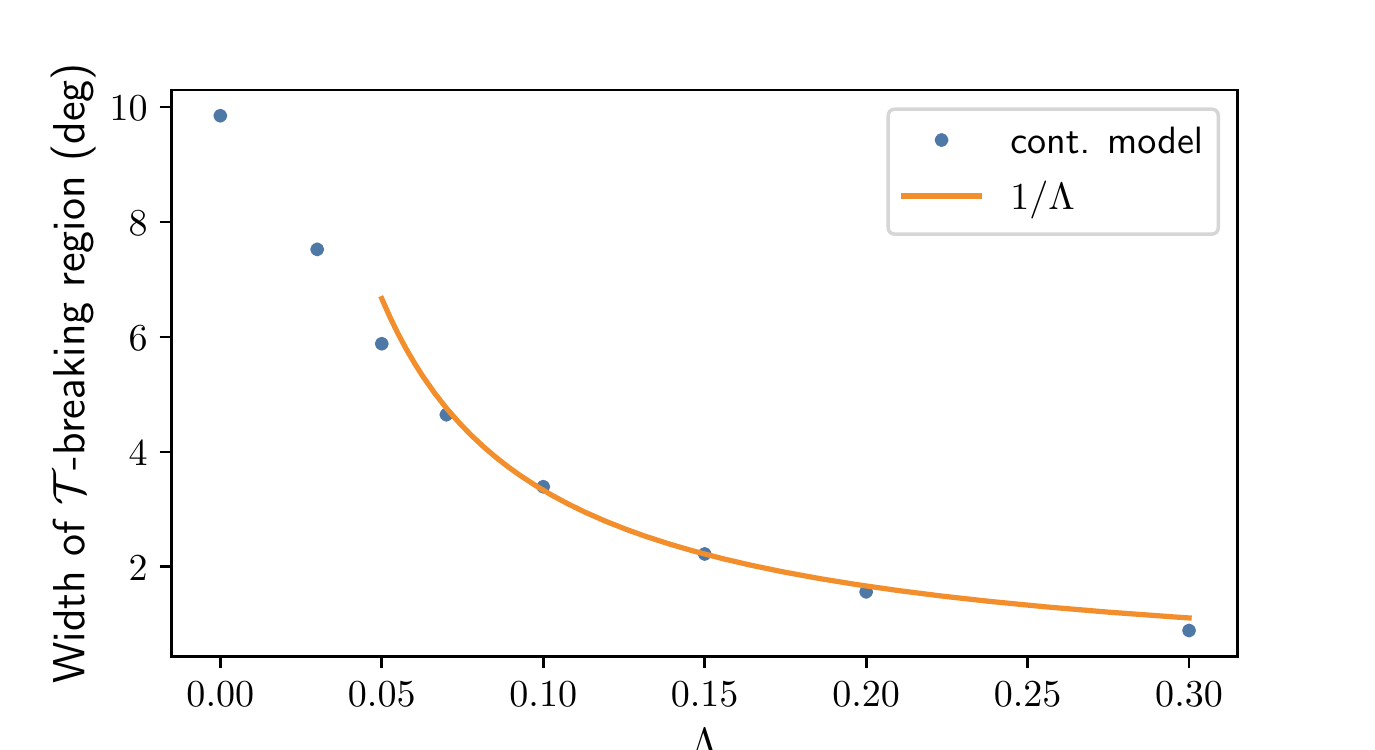}
    \caption{$\Lambda$-dependence of the width of the $\mathcal{T}$-breaking region at $T=0$ computed in the continuum model. It is well approximated by a $1/\Lambda$ (orange line) for large $\Lambda$.}
    \label{fig:lambdascaling2}
\end{figure}

As discussed in Sec.~\ref{sec:freeenergy}, the width of the topological phase space region along $\theta$ is
related to $J_{c2}(\theta=0)/J_{c1}(\theta=0) \sim 1/\Lambda$ which we numerically confirm in Fig.~\ref{fig:lambdascaling2}.

\section{Numerical evaluation of crossed diagram}
\label{apdx:complexity}

The crossed diagram in Fig~\ref{fig:diagrams}(c) involves evaluation of the sum
\begin{align}
    \sum_{n\mathbf{kqq'}} 
	\left|\gamma_\mathbf{q} \right|^2
        \left|\gamma_\mathbf{q'} \right|^2
         f_{n\mathbf{k},1}f_{n\mathbf{k+q+q'},1}
        f_{n\mathbf{k+q},2}f_{n\mathbf{k+q'},2} \,.
        \label{eqn:N1}
\end{align}
Naively, since this sum does not factorize and convolutional structure is not apparent, numerical evaluation requires computational time $\mathcal{O}(MN^3)$, where $M$ is the number of Matsubara frequencies required to reach convergence and $N$ are the number of $\mathbf{k}$-points in the 2D Brillouin zone. Here, we show that the complexity can be reduced to $\mathcal{O}(MN^2)$ by exploiting the Gaussian form of $\left|\gamma_\mathbf{q} \right|^2 \sim e^{-\mathbf{q^2}/\Lambda^2}$. For ease of notation, we will omit prefactors and Matsubara indices $n$ and set $\Lambda =1$. 

We perform the substitution $\mathbf{q}\rightarrow \mathbf{q-k}$ and $\mathbf{q'}\rightarrow \mathbf{q'-k}$ which modifies Eq.~\eqref{eqn:N1} to
\begin{align}
    &\sum_{\mathbf{qq'}}
    e^{-(\mathbf{q-q'})^2/2} f_{\mathbf{q},2} f_{\mathbf{q'},2}
    \\
    &\times
    \sum_{\mathbf{k}}
    \exp \left( -\frac{1}{2} (2\mathbf{k}-\mathbf{q}-\mathbf{q'})^2  \right)
    f_{\mathbf{k},1}f_{\mathbf{q+q'-k},1}.
\end{align}
Choosing the center of mass frame $\mathbf{Q}=\mathbf{q+q'}$, $\mathbf{P}=\mathbf{q-q'}$, the term is expressed as
\begin{align}
    &\frac{1}{4}\sum_{\mathbf{P}} e^{-\mathbf{P}^2/2} \sum_{\mathbf{Q}} f_{\frac{\mathbf{P+Q}}{2},2} f_{\frac{\mathbf{P-Q}}{2},2}
    \nonumber
    \\
    &\times
    \sum_{\mathbf{k}}
    \exp \left( -\frac{1}{2} (2\mathbf{k}-\mathbf{Q})^2  \right)
    f_{\mathbf{k},1}f_{\mathbf{Q-k},1}.
\end{align}
Them sum over $\mathbf{k}$ requires $N^2$ steps ($N$ steps to perform the $\mathbf{k}$-summation times $N$ steps to resolve the residual $\mathbf{Q}$-dependence). Analogously, the $\mathbf{Q}$-sum can be computed in $N^2$ steps and the remaining $\mathbf{P}$-sum is performed in $N$ steps. Taking into account the Matsubara summation, we arrive at a total complexity of $\mathcal{O}(MN^2)$.


\section{Surface spectral function}
\label{apdx:surfacegf}

In the following we summarize the derivation of the surface Green's function given in Ref.\,\cite{Pinon.2020}. This method provides a remarkably clear and numerically efficient way to examine the boundary spectrum, given a bulk Green's function $G(\mathbf{k})$. It involves perturbing the original Hamiltonian with an `impurity' line of the form
\begin{align}
	\mathcal{U} = u_0\sum_{\mathbf{r}}^{}\delta_{x,0}  
	c_{\mathbf{r}}^\dagger  c_{\mathbf{r}} \,.
\end{align}
Here, the coordinate vector is $\mathbf{r}=(x,y)$. We see that the perturbation introduces a potential barrier along the line $x=0$. If the barrier is sufficiently high, i.e. $u_0 $ is much larger than the bandwidth, tunnelling between the two infinite half planes $x>0$ and $x<0$ will be completely suppressed and they become essentially decoupled. In effect, the perturbation creates two independent semi-infinite half planes with boundaries at $x=\pm1$.

The perturbation $\mathcal{U}$ can be treated exactly by means of a Dyson series for the full Green's function $F(\mathbf{p};\mathbf{p}')$
\begin{align}
	F(\mathbf{p};\mathbf{p}') =
	G(\mathbf{p})\delta_{\mathbf{p,p'}} 
    +
	G(\mathbf{p})
	\sum_{\mathbf{q}}
	U(\mathbf{p-q}) F(\mathbf{q};\mathbf{p'}),
\end{align}
where we suppress the frequency variable for brevity. Here, $G(\mathbf{p})$ is the disorder-averaged Green's function defined in Eq.~\eqref{eq:gfd} and $U(\mathbf{q}) = u_0\sigma_z\delta_{q_y,0}$ is the first-quantized matrix of the perturbation $\mathcal{U}$. Note that $F(\mathbf{p};\mathbf{p}')$ is not diagonal in momentum space, since translational invariance is explicitly broken by the perturbation. We rewrite the above series using the transfer matrix $T$, yielding
\begin{align}
	F(\mathbf{p;p'}) = 
	G(\mathbf{p})\delta_{\mathbf{p,p'}} 
	+
	G(\mathbf{p})
	T(\mathbf{p;p'})
	G(\mathbf{p'})
	\label{eqn:Ffull}
\end{align}
with an explicit expression of the transfer matrix
\begin{align}
	T(\mathbf{p;p'}) &=
	\left[\mathbb{1}-u_0\sigma_z\sum_{q_x}^{} G(q_{x}p_y)\right]^{-1} u_0\sigma_z
	\delta_{p_y,p_y'} \,.
	\label{eqn:t2}
\end{align}
The local Green's function $\tilde{G}$ is deduced by transforming the $p_x$ momentum coordinate to real space:
\begin{align}
	\tilde{G}(x,p_y) &\equiv F(xp_y;xp_y) \nonumber \\
    &= \sum_{p_xp_x'}
	e^{ip_x x}e^{-ip_x' x} F(p_x p_y;p_x' p_y) \,.
\end{align}
The surface Green's function is obtained by evaluating $\tilde{G}(x,p_y)$ at $x=\pm1$ and the corresponding spectral function is plotted in Fig.~\ref{fig:edgemode}.

\begin{figure}[t]
	\centering
	\includegraphics[width=0.8\columnwidth]{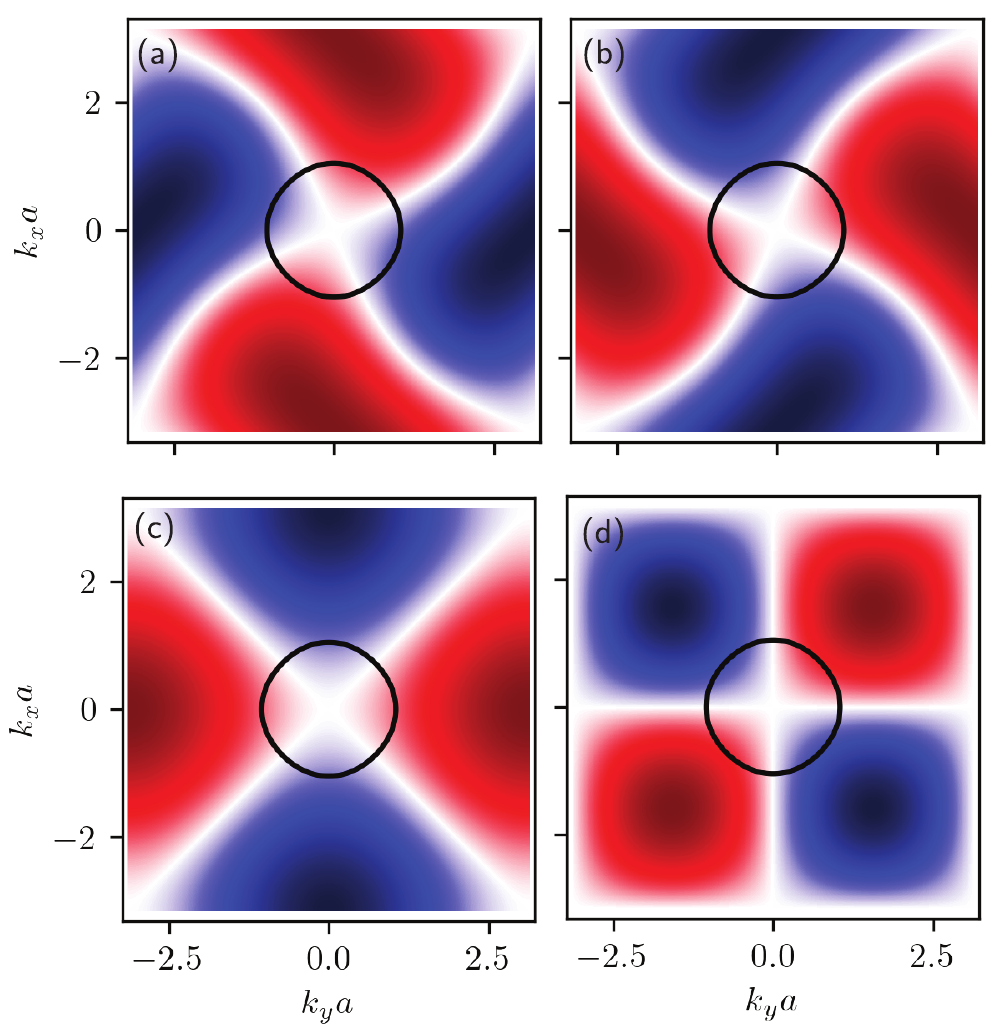}
	\caption{Two different gap regularizations. Panels (a-b) show the gaps defined in Eq.~\eqref{eq:alignedgap}. They can be transformed into each other by a mirror reflection, up to a sign. Panels (c-d) show a different regularization where this symmetry is absent.}
	\label{fig:gaps}
\end{figure}


\section{Lattice regularizations of $\Delta_{\mathbf{k},i}$}

\begin{figure}[ht]
	\centering
	\includegraphics[width=\columnwidth]{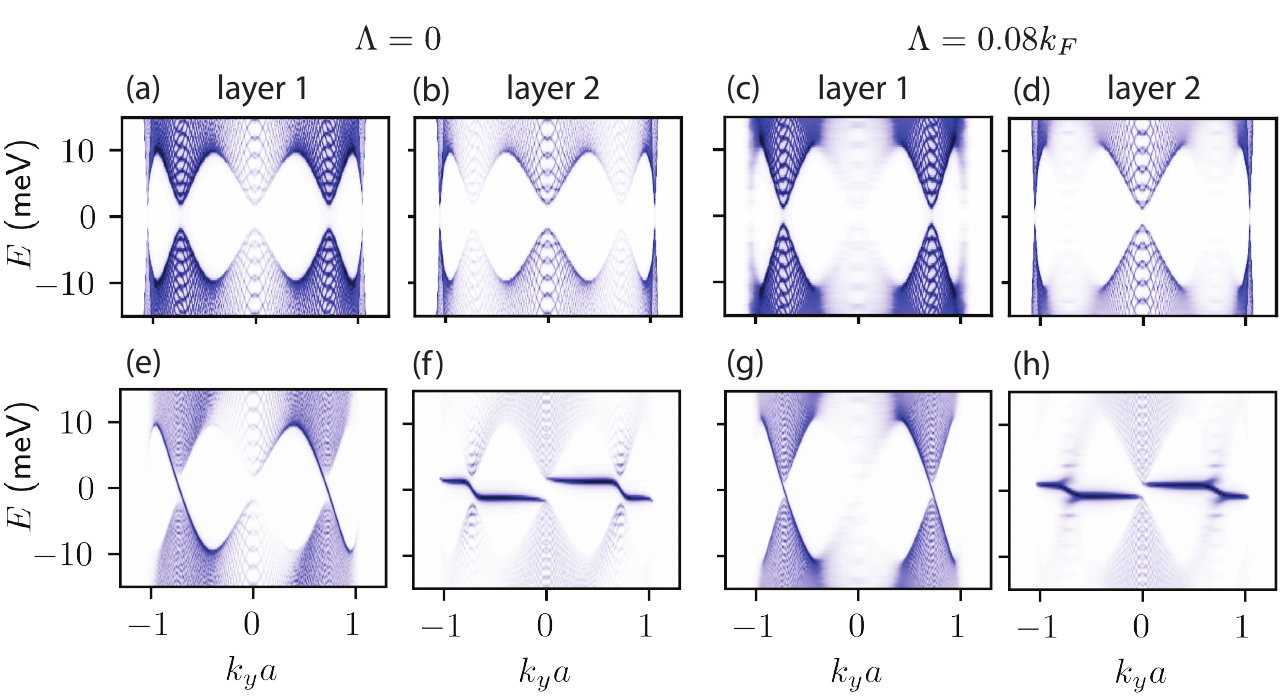}
	\caption{Layer-resolved bulk (a-d) and boundary (e-h) spectra for incoherently coupled cuprate bilayers with $\Lambda=0$ (left two columns) and $\Lambda/k_F=0.08$ (right two columns) at $45^\circ$ twist angle for superconducting gaps plotted in Fig.~\ref{fig:gaps}(c-d). In each layer, the spectral functions display two distinct chiral edge modes, indicating a total Chern number $\mathcal{C}=4$.}
	\label{fig:edgemode2}
\end{figure}

Regularization of the superconducting gap function in Eq.~\eqref{eq:alignedgap} gives rise to an accidental degeneracy of the surface spectral function in Fig.~\ref{fig:edgemode}. This is because the gap functions in each layer can be mirrored into each other with respect to the axis $k_x=0$, up to a phase of $-1$. Plots of the two gaps are shown in the top panels of Fig.~\ref{fig:gaps} where the Fermi surface is indicated by a black circle. When projected onto the $k_y$-axis, the two gaps will yield identical spectra.

A different regularization, specific to the twist $\theta=\pf$, is given by
\begin{align}
    \Delta_{\mathbf{k},1} &= \Delta\sin k_x  \sin k_y, \nonumber \\
    \Delta_{\mathbf{k},2} &= \Delta \left( \cos k_x - \cos k_y \right) \,.
\end{align}
The two gap functions are plotted in the bottom panels of Fig.~\ref{fig:gaps}. Here, the two gaps are no longer related to each other through a mirror symmetry. The corresponding layer-resolved spectral functions hence differ in the two layers of the bilayer structure, as seen in Fig.~\ref{fig:edgemode2}. In the absence of any interlayer coupling, $g=0$, each layer possess four Dirac cones. Finite interlayer coupling $g$ has two effects: (a) it gaps the Dirac cones and (b) induces the four gapped cones of layer $1$ onto layer $2$ and vice versa. The induced gapped Dirac cones are characterized by light spectral weight in Fig.~\ref{fig:edgemode2}. The positions of gapped Dirac cones correspond to intersections of the Fermi surface in Fig.~\ref{fig:gaps}(c-d) with the gap nodes, projected onto the $k_y$ axis.

In each layer, the gapped Dirac cones are traversed by two Chiral edge modes, indicating a total topological Chern invariant of $\mathcal{C}=4$. This is still the case in the presence of impurities, $\Lambda/k_F=0.08$. Here, the incoherent nature of the interlayer tunneling causes a broadening of the induced gapped Dirac cones.


\end{document}